\documentclass[12pt]{article}
\usepackage{epsfig}
\usepackage{amsfonts} 
\usepackage{amssymb}  
\usepackage{graphicx} 
\usepackage{amsmath}  
\usepackage{amsmath}  
\usepackage[latin1]{inputenc}
\usepackage{color}
\def\Title#1{\begin{center} {\Large {\bf #1} } \end{center}}

\begin{document}

\Title{Anomalous-Magnetic-Moment Effects in a Strongly Magnetized and Dense Medium}

\bigskip\bigskip


\begin{raggedright}
{\it E. J. Ferrer\index{Ferrer, E}\\
Department of Physics, University of Texas at El Paso,
El Paso, TX 79968, USA\\
 Email: ejferrer@utep.edu\\
V de la Incera \index{Incera, V}\\
Department of Physics, University of Texas at El Paso,
El Paso, TX 79968, USA\\
Email: vincera@utep.edu\\
 D. Manreza Paret\index{Manreza, D}\\
Departamento de Fisica General, Facultad de Fisica, Universidad de la Habana, La Habana, 10400, Cuba\\
Email: dmanreza@fisica.uh.cu\\
 A. P\'{e}rez Mart\'{\i}nez\index{Perez Martinez, A}\\
Instituto de Cibern\'{e}tica, Matem\'{a}tica y F\'{\i}sica (ICIMAF),
 La Habana, 10400, Cuba\\
Email: aurora@icimaf.cu
}
\bigskip\bigskip
\end{raggedright}

\section{Introduction}

A main problem in astrophysics is to establish a connection between the astrophysical observations of neutron stars and the composition and state of the matter that forms them. A way to discriminate different proposals for the inner star composition is to study the equation of state (EoS) of different models of nuclear and quark matter, obtain the curves of mass versus radius coming from those EoS's,
 and compare them  with observed mass-radii of the
neutron stars.  The EoS is configured from the star inner content and external conditions as density, temperature, fields, etc. The fact that strong magnetic fields populate the vast majority of the astrophysical compact objects and that they can significantly affect several properties of the star, have served as motivation for many works focused on the study of the EoS's of magnetized systems of fermions \cite{MNS, EoS-AMM1, EoS-AMM2} and their astrophysical implications.

A direct consequence of an external magnetic field, $B$, is the modification of the fermion energy dispersion $E=\sqrt{p^2_3+2eBl+m^2}$ due to the Landau level (LL) quantization \cite{Landau} of the momentum component transverse to the magnetic field. An external magnetic field also affects the density of states which now becomes proportional to the field. Hence,
in the presence of a uniform and constant magnetic field one should do the replacement
\begin{equation}
2\int\frac{d^3p}{(2\pi)^3}\rightarrow\sum_l g(l)\frac{eB}{(2\pi)^2}\int dp_3
\end{equation}\label{density_state}
\noindent in all the integrals of the momenta, which corresponds to the charged particles (as we will see, for instance in Eq (\ref{Grand-Potential-4})). The factor $g(l)=[2-(\delta_{l0})]$ takes into account the double spin degeneracy of all the Landau levels except $l=0$.

The Dirac Hamiltonian of the theory can be modified by considering the radiative corrections that come from the fermion self-energy.  The one-loop self-energy of a charged fermion in a magnetic field has a term of Dirac structure $\gamma_1\gamma_2$, that corresponds to the fermion anomalous magnetic moment (AMM) $\tau$.  When the radiative corrections are included in the effective Hamiltonian of the theory, the AMM  gives rise to an interaction of the form $\frac{1}{2}\tau \sigma_{\mu\nu}F^{\mu\nu}$. This new term in the Hamiltonian accounts for the coupling between the external magnetic field and the AMM.

At weak fields ($eB\ll m^2$), one can make an expansion in powers of the magnetic field and find that the leading contribution to the coefficient of the structure $\gamma_1\gamma_2$ in the self-energy is given (in natural units) by $\tau=(\alpha/2\pi)(e/2m)=(\alpha/2\pi)\mu_B$, as shown by Schwinger \cite{Schwinger} many years ago. So at weak fields $\tau$ is simply independent of the magnetic field.  At finite temperature and/or density, the concept of weak field needs to be revisited, as the field may be weak with respect to one of the scales, but strong with respect to the
others. The inclusion of the AMM in the Dirac Hamiltonian breaks the spin degeneracy for $l >0$.

The AMM term in the Hamiltonian changes the energy spectrum of the fermions and can affect in principle the properties of the system. Notice that certain neutral particles which, like the neutron, have an internal structure, can also have nonzero AMM. The effects of the nucleons' and quarks' AMM on the statistics of magnetized matter have been discussed in many works \cite{EoS-AMM1, EoS-AMM2}. The AMM has been linked among other effects to stiffening the EoS in magnetized stars and to a dramatic variation of the proton fraction, which at very high magnetic fields would lead to pure neutron matter \cite{EoS-AMM1}.

In the present work we are interested in the effects of the AMM on a system of fermions for both
 weak and strong magnetic fields. Notice that in the strong field region the Schwinger approximation to the AMM is not valid anymore, and any conclusion drawn from using the linear approximation in that region would be inconsistent.  This issue has been already pointed out many years ago in Ref. \cite{Jancovinci}.
In what follows we shall analyze, through analytical and numerical calculations, the significance of the AMM contribution to the main statistical quantities of the magnetized system, as well as to the EoS, in the weak and strong field approximations.

The paper is organized as follow. In Sec. \ref{section2} we give the one-loop self-energy of a charged fermion system in the presence of a constant and uniform magnetic field using the Ritus's method \cite{Ritus:1978cj}, and find the AMM analytical expression in the strong-field limit. In Sec. \ref{section3}, the dispersion relation is obtained taking into account the AMM correction, and the results at strong and weak fields are compared. In Sec. \ref{section4} we present the thermodynamical potential including the AMM for each Landau level, study the corresponding EoS in the strong field limit, and compare the results with those obtained either at zero AMM or with the Schwinger approximation for the AMM. In Sec. \ref{section5}, we present the numerical results in the weak- and strong-magnetic-filed limits, for the main thermodynamic quantities, which depend on the AMM. Finally, in Sec. 6 we state our concluding remarks.


\section{Landau Level Dependence of the AMM}\label{section2}

In this section we present the calculation of the AMM contribution to the one-loop fermion self energy, $\Sigma^{l}(\overline{p})$, in the presence of a constant and uniform magnetic field. The general structure of the self energy in momentum space is \cite{Incera}
\begin{equation}\label{SE-structure}
\Sigma^{l}(\overline{p})
=Z_{\|}^{l}\overline{p}_{\|}^\mu\gamma_{\mu}^{\|}+Z_{\bot}^{l}\overline{p}_{\bot}^\mu\gamma_{\mu}^{\bot}+M^{l}I+iT^{l}\gamma^{1}\gamma^{2},
\end{equation}
where the separation between parallel ${\overline{p}}_{\|}^\nu=(p^{0},0, 0,p^{3})$ and perpendicular ${\overline{p}}_{\bot}^\nu=(0,0, \sqrt{2eBl},0)$ components of the fourth momentum is a direct consequence of the explicit breaking of the rotational symmetry by the external magnetic field. Only the subgroup of rotations about the direction of the field remains intact. In (\ref{SE-structure}), $Z_{\|}^{l}$, $Z_{\bot}^{l}$ are the wave function's renormalization coefficients. The coefficients $M^{l}$ and $T^l$ corresponds to the mass and anomalous magnetic moment respectively. Each of them have to be determined as solutions of the Schwinger-Dyson (SD) equations of the theory at the given approximation. We will find them in the one-loop approximation.

Using Ritus' approach one can show that the  Schwinger-Dyson system of equations for all Landau level numbers $l's$ takes the form  \cite{VyE}
\begin{eqnarray}
\Sigma^{l}(\overline{p})\Pi(l)&=&-ie^2(2eB)\Pi(l)\int\frac{d^4\widehat{q}}{(2\pi)^4}
\frac{e^{-\widehat{q}^2_\bot}}{\widehat{q}^2}[L_l+L_{l+1}+L_{l-1}], \quad l=0,1,2,....\label{SD-EqL}
\end{eqnarray}
where the $L$ factors are given by
\begin{eqnarray}
  L_l &=& \gamma_{\mu}^{\|}G^{l}(\overline{p-q})\gamma_{\mu}^{\|}, \nonumber\\
  L_{l\pm1} &=& \Delta(\pm)\gamma_{\mu}^{\bot}G^{l\pm1}(\overline{p-q})\gamma_{\mu}^{\bot}\Delta(\pm)
 \label{expresiones1}
\end{eqnarray}
and the fermion propagator is $G^l(\overline{p})=-\dfrac{\overline{p}\cdot\gamma+m}{\overline{p}^2+m^2}$. Here we introduced the spin projectors $\Delta(\pm)=\frac{I\pm i\gamma^{1}\gamma^{2}}{2}$; a spin degeneracy factor $\Pi(l) = \Delta(+)\delta^{l0}+I(1-\delta^{l0})$ that separates the lowest landau level (LLL) from the rest; and used the notation $\widehat{q}_\mu=q_\mu/\sqrt{2eB}$ and $(\overline{p-q})_l = (p^{0}-q^{0},0,-\sqrt{2eBl},p^{3}-q^{3})$. Without loss of generality, we can assume that $eB>0$. From Eq. (\ref{SD-EqL}) we can extract the equations for the AMM at each LL. In Euclidean space they are
\begin{eqnarray}
(M^0+T^0)=e^2(2eB)\int\frac{d^4\widehat{q}}{(2\pi)^4}
\frac{e^{-\widehat{q}^2_\bot}}{\widehat{q}^2}\left(\frac{2m}{(\overline{p-q})^2_0+m^2} +\frac{2m}{(\overline{p-q})^2_{1}+m^2}\right ),\label{LLLMT}
\end{eqnarray}

\begin{eqnarray}
  T^{l} &=&-e^2m(2eB)\int\frac{d^4\widehat{q}}{(2\pi)^4}\frac{e^{-\widehat{q}^2_\bot}}{\widehat{q}^2}.
  \left[\frac{1}{(\overline{p-q})^2_{l+1}+m^2}-\frac{1}{(\overline{p-q})^2_{l-1}+m^2} \right] \label{generalLL}
\end{eqnarray}
Note  that Eq. (\ref{LLLMT}) reflects  the fact that fermions in the LLL ($l=0$) have only one spin orientation, and as a consequence, it is impossible to determine $M^0$ and $T^0$ independently \cite{VyE, Vivian}. The representation of the self-energy operator (\ref{SD-EqL}) is particularly convenient for calculations in the strong-field approximation, where the LLL gives the leading contribution.

Considering the strong field case and taking the infrared limit ($p_0=0, p_3\rightarrow 0$) in (\ref{LLLMT}), we obtain
\begin{equation}
  E^0=M^{0}+T^{0} = \frac{e^2m}{8\pi^2}\int d\widehat{q}_{\|}^2d\widehat{q}_{\bot}^2\frac{e^{-\widehat{q}^2_\bot}}{\widehat{q}^2}  \left[\frac{1}{\widehat{q}_{\|}^2+\widehat{m}^2}+\frac{1}{\widehat{q}_{\|}^2+1+\widehat{m}^2} \right],\label{LLLMT1}
\end{equation}
Integrating in $\widehat{q}_{\|}^2$ and $\widehat{q}_{\bot}^2$ we have
\begin{eqnarray} \label{LLLMT2}
  E^{0}&= &\frac{e^2m}{8\pi^2}\int d\widehat{q}_{\bot}^2e^{-\widehat{q}^2_\bot}  \left[\frac{\ln\frac{\widehat{q}_{\bot}^2}{\widehat{m}^2}}{\widehat{q}_{\bot}^2-\widehat{m}^2}
  +\frac{\ln\frac{\widehat{q}_{\bot}^2}{\widehat{m}^2+1}}{\widehat{q}_{\bot}^2-\widehat{m}^2-1} \right]  \simeq\frac{\alpha}{4\pi}\log^2(\widehat{m}^2)
\end{eqnarray}

Expression (\ref{LLLMT2}) gives the LLL AMM contribution to the self-energy in the strong-field limit. In contrast to the Schwinger AMM term \cite{Schwinger}, which grows linearly with the magnetic field and affects all the LLs in the same way, at strong field, only the LLL gets a nonzero AMM, which grows as a square logarithm of the field. Clearly, using the Schwinger term in the strong-field region would be totally inconsistent and care should be taken not to draw any physical conclusions obtained with such a wrong approach.

\section{Dispersion Relations}\label{section3}

The fermion dispersion relations including the AMM through radiative corrections are obtained from
\begin{equation}
detG^{-1}_l=det[\overline{p}\cdot\gamma-m-M^lI-iT^l\gamma^1\gamma^2]=0,
\end{equation}\label{Inverse-Prop}
where the inverse fermion propagator $G^{-1}_l$ incorporates the one-loop radiative corrections to the Dirac Hamiltonian, which includes the AMM contribution $T^l$.
The dispersion relations are then
\begin{equation}\label{Disp-Rel-1}
\epsilon_{\sigma, l}^2=p_3^2+[\sqrt{(m +M^l)^2 +2|eB|l}+\sigma T^l)]^2,  \quad \sigma=\pm1
\end{equation}
for $l\geq 0$  and
\begin{equation}\label{Disp-Rel-3}
\epsilon_{1,0}^2=p_3^2+(m+E^0)^2
\end{equation}
for $l=0$.

Considering the Schwinger AMM,  the dispersion relation becomes
\begin{equation}
detG^{-1}_l=det[\overline{p}\cdot\gamma-m-\kappa\mu_BB\sigma]=0,
\end{equation}\label{Inverse-Prop1}
and the corresponding energy spectrum
\begin{equation}
 \epsilon_{\sigma,l}^2=p_3^2 + [(m^2+2eBl)^{1/2}-\kappa\mu_BB\sigma]^2  \label{Sch}
\end{equation}
is the same for all LL's.
Hence, the particle rest-energy in the LLL is
\begin{align}
\varepsilon^{0}_{\text{\tiny Sch}} = \mid m -\kappa\mu_B B\mid,\label{energy1}
\end{align}
From Eq. (\ref{energy1}) it seems that for sufficiently strong magnetic fields the rest-energy vanishes. This result is a consequence to extrapolate the Schwinger AMM term to high magnetic field without taking into account that the Schwinger derivation is only valid at weak field where the linear approximation becomes the leading one.

\begin{figure}[ht!]
\begin{center}
\includegraphics[width=10cm,height=8cm]{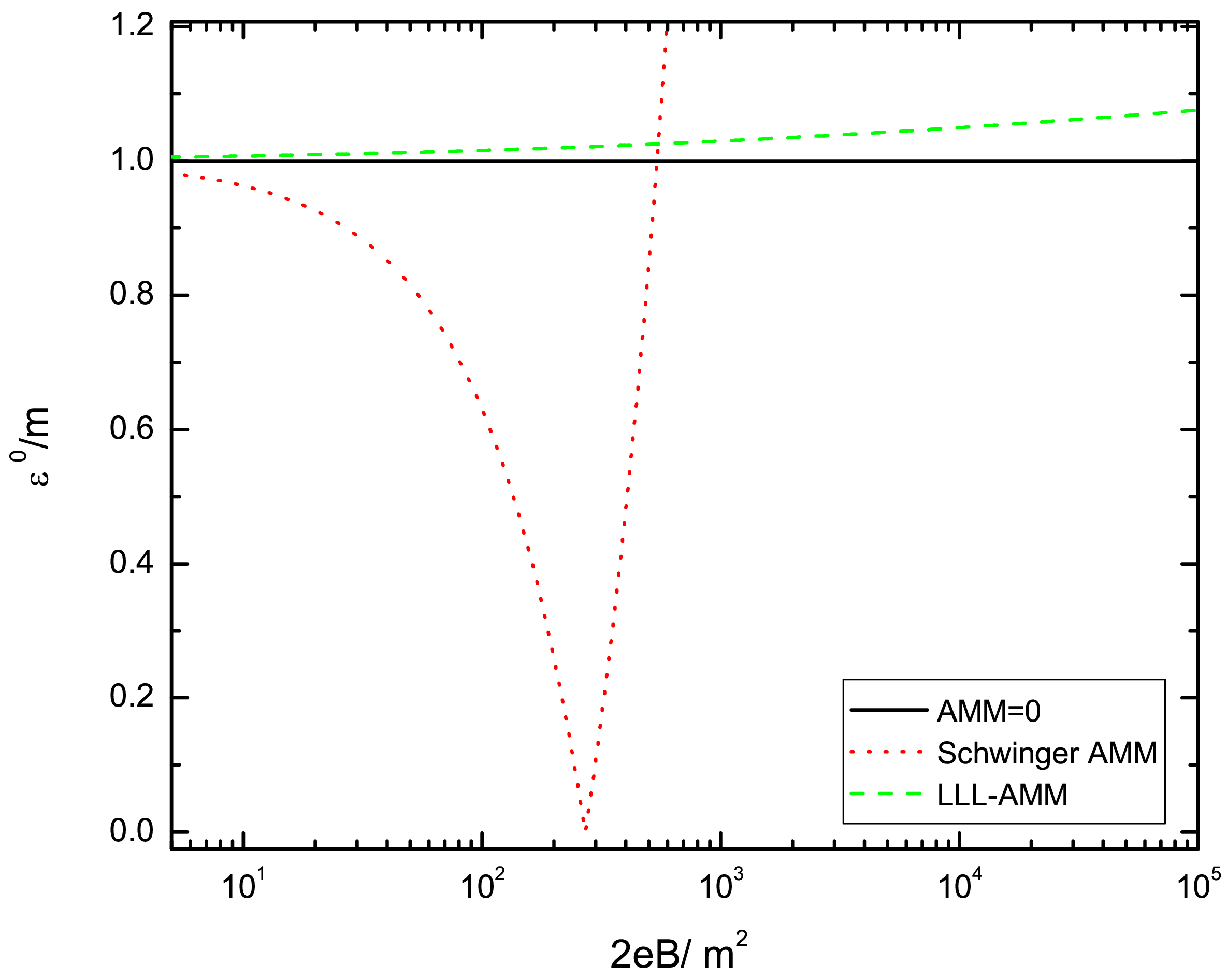}
\caption{\footnotesize Rest energy of the particles considering AMM $(\varepsilon^{0}_{1,0})$ and Schwinger approximation $(\varepsilon^{0}_{\text{\tiny Sch}})$ versus $2eB/m^2$ for strong field values. The case of no AMM $(\varepsilon^0)$ is also shown for comparison.}
\label{fig1}
\end{center}
\end{figure}

Thus, we conclude that at strong fields the particle rest energy is given by
\begin{equation}\label{Disp-Rel-4}
\varepsilon^{0}_{1,0} =\mid m+E^0\mid.
\end{equation}

In Fig (\ref{fig1}) we plotted the rest energies  $\varepsilon^{0}_{1,0} $ and $\varepsilon^{0}_{\text{\tiny Sch}}$ versus $2eB/m^2$ for strong field values. From the plots we see that the behaviors of $\varepsilon^{0}_{\text{\tiny Sch}}$ and $\varepsilon^{0}_{1,0} $ are totally different. While $\varepsilon^{0}_{\text{\tiny Sch}}$ decreases with the field, reaching zero value for magnetic field strengths approximately three order larger than the mass square, the rest energy $\varepsilon^{0}_{1,0} $ has a moderate increase with respect to that with zero AMM (i.e. $\varepsilon^0=m$).

Considering the Schwinger AMM in QED we find that the threshold field for zero rest energy is $B\simeq 10^{15}$ G, while in QCD, considering the u-quark mass, the threshold field is of order $B\simeq 10^{18}$ G. But as discussed above, the Schwinger approximation is only valid for $eB\ll m^2$, and what is seen from our result, $\varepsilon^{0}_{1,0} $, is that the effect of the AMM at strong field is to increase the particle rest energy instead of decreasing it.

\section{Thermodynamical Potential with AMM in the Strong Field Limit}\label{section4}

In this section we are going to explore the impact of the AMM in the thermodynamical quantities of the magnetized fermion system. With this aim we start from the fermion contribution to the thermodynamical potential ($\Omega_f$) at finite temperature and density, and in the presence of a uniform and constant magnetic field, but including the AMM quantum correction in the fermion inverse propagator.
\begin{equation}\label{Grand-Potential-4}
\Omega_f= -\frac{eB}{\beta}\left[\sum_{p_4}\int\limits_{-\infty}^{\infty}\frac{dp_3}{(2\pi)^2} \ln \det G^{-1}_0(\overline{p}^*)+ \sum_{\sigma\pm 1}\sum_{l=1}^{\infty}\sum_{p_4}\int\limits_{-\infty}^{\infty}\frac{dp_3}{(2\pi)^2} \ln \det G^{-1}_l(\overline{p}^*)\right],
\end{equation}
with ${\overline{p}}^{*}=(ip^{4}-\mu,0, \sqrt{2eBl},p^{3})$ for $l=0,1,2,...$;  $\beta$ is the inverse absolute temperature, $\mu$ is the fermionic chemical potential, and we separated the LLL contribution from the rest.

After doing the Matsubara sum and taking the zero temperature limit, we can write the thermodynamic potential as a sum of the vacuum and statistical contributions
\begin{equation}\label{Omega1}
  \Omega_f= \Omega_f(B,0,0) + \Omega_f(B,\mu,0).
\end{equation}
with
\begin{equation}\label{Omegav}
 \Omega_f(B,0,0)= -\frac{eB}{4\pi^2}\int dp_3|\varepsilon_{1,0}|-\frac{eB}{4\pi^2}\sum_{\sigma}\sum_{l=0}^{\infty}\int dp_3|\varepsilon_{\sigma,l}|,
\end{equation}
and
\begin{equation}\label{OM}
   \Omega_f(B,\mu,0) =-\frac{eB}{4\pi ^2}\left [\Omega_{LLL} + \sum_{l=1}^{l_{max}}\sum_{\sigma=\pm 1}\left( \mu\,{p}_{F}^{\sigma} -  ({\varepsilon}_{\sigma,l}^{0})^2\ln\frac{ {\mu} + {p}^{\sigma}_{F}}{{\varepsilon}_{\sigma,l}^{0}}\right)\right ],
\end{equation}
where $l_{max}= [\frac{(\mu-\sigma T^l)^2-(m+M^l)^2}{2eB}]$, $I[z]$ denotes the integer part of $z$,
\begin{equation}\label{Om_mag0}
 {\Omega}_{LLL} =\left ( {\mu}\,{p}_{F}^{0} -  ({\varepsilon}_{1,0}^{0})^2
\ln\frac{  {\mu} +{p}_{F}^0}{{\varepsilon}_{1,0}^0}\right ),
\end{equation}
and the Fermi momenta for zero and nonzero LL are respectively ${p}_F^{0}=\sqrt{{\mu}^2-({\varepsilon}_{1,0}^{0})^2}$ and ${p}_{F}^{\sigma}=\sqrt{\mu^2-({\varepsilon}_{\sigma,l}^{0})^2}$.
Here we used the notation $({\varepsilon}_{1,0}^{0})^2=m^2 +(E^0)^2$ and $(\varepsilon_{1,l}^{0})^2=(\sqrt{(m +M^l)^2 +2|eB|l}+\sigma T^l)^2$.

Henceforth we will concentrate in the strong-field region, which is where the effect of the AMM is more significant. That means that we will assume that $eB\gg \mu^2 \gg m^2$. In this case, we can neglect the radiative corrections in all the LLs except the lowest one, so in Eqs.(\ref {Omegav})-(\ref{Om_mag0}) we can replace $\varepsilon^0_{\sigma,l}$ by $\varepsilon^0_{l}= \sqrt{2|eB|l +m^2}$ for the two spin projections. In addition, we can drop the contributions of all the LLs except $l=0$ because in this approximation $l_{max}=0$. All these considerations significantly simplify our calculations.

The vacuum contribution $\Omega_f(B,0,0)$ needs to be renormalized. This can be done by adding and subtracting the contribution of the LLL with $E^0 =0$ and following the Schwinger's procedure \cite{Proper-Time} to renormalize the part that corresponds to the usual vacuum term. In this way, we obtain the renormalized vacuum contribution
\begin{equation}\label{vacuum-strong}
\Omega_f^R(B,0,0)=-\frac{eB}{4\pi^2}\int_0^{\infty} dp_3 \left[|\varepsilon_{1,0}|- \sqrt{p_3^2+m^2}\right ]+\Omega_{AMM=0}^R(B,0,0),
\end{equation}
where $\Omega_{AMM=0}^R(B,0,0)$ is the well-known renormalized thermodynamic potential in the strong field limit and at zero AMM \cite{Ragazzon}.
\begin{equation}\label{Omega-renormalized}
   \Omega_{AMM=0}^R(B,0,0)=-\frac{\alpha}{6\pi}B^2\ln \left (\frac{eB}{m^2}\right )
\end{equation}

We can work with the first term in (\ref{vacuum-strong})  to write it in the proper-time form, and then put together all the contributions that remain in the strong-field limit to find

\begin{equation}\label{Omega-strongB}
   \overline{\Omega}_f^R=\frac{\alpha \overline{B}^2}{4\pi}\int_1^{\infty} \frac{ds}{s^2} \left[e^{-s\frac{(\overline{\varepsilon}^0_{1,0})^2}{e\overline{B}}}- e^{-s\frac{\overline{m}^2}{e\overline{B}}}\right ]+\overline{\Omega}_{AMM=0}^R-\frac{e\overline{B}}{4\pi^2} \overline{\Omega}_{LLL}.
\end{equation}

In the above expression, for convenience and as a preparation for the numerical calculations, we normalized all the quantities with respect to the chemical potential.

\section{Numerical Results}\label{section5}

We are now ready to numerically find the strong-field behavior of the thermodynamical functions in the effective theory described by the thermodynamic potential (\ref{Omega-strongB}). These quantities are
the magnetization $ \mathcal{M}=-(\partial \Omega_f^R/\partial B)$, the energy density $\epsilon= B^2/2+\Omega_f^R+ \mu N$, and the parallel $P_\parallel = -\Omega_f^R-B^2/2$ and transverse $P_\bot = -\Omega_f^R-B\mathcal{M}+B^2/2$ pressures of the system. Notice that the pure Maxwell contribution $B^2/2$ should be added to find the energy and pressures of the system \cite{PRC82}.
\begin{figure}[!ht]
\begin{center}
\includegraphics[width=8cm,height=6cm]{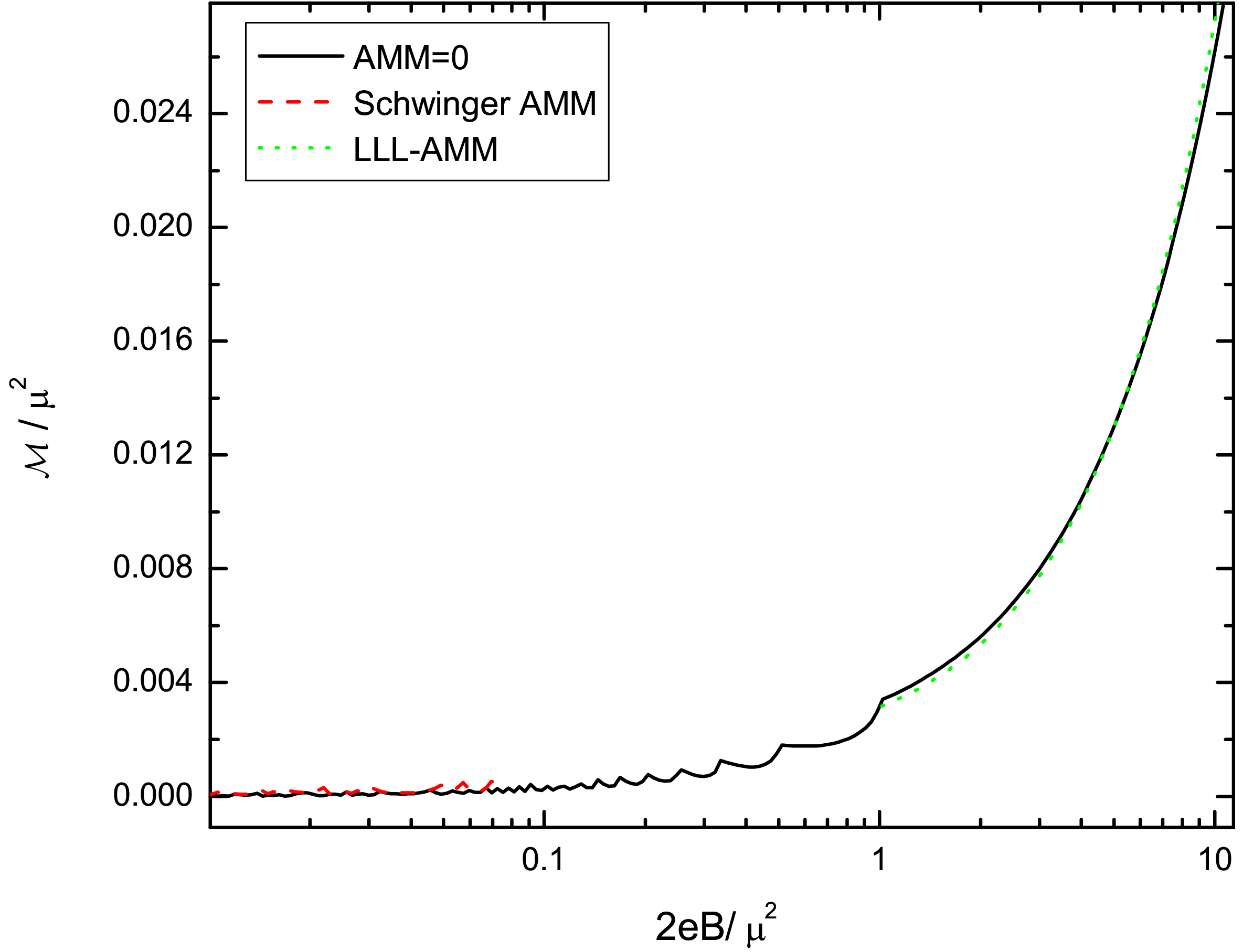}
\caption{\footnotesize Magnetization versus $2eB/\mu^2$ with $m/\mu\simeq 0.02$. Comparison between AMM in LLL, Schwinger approximation of AMM, and  without AMM cases.} \label{fig2}
\end{center}
\end{figure}
\begin{figure}[!ht]
\begin{center}
\includegraphics[width=14cm,height=6cm]{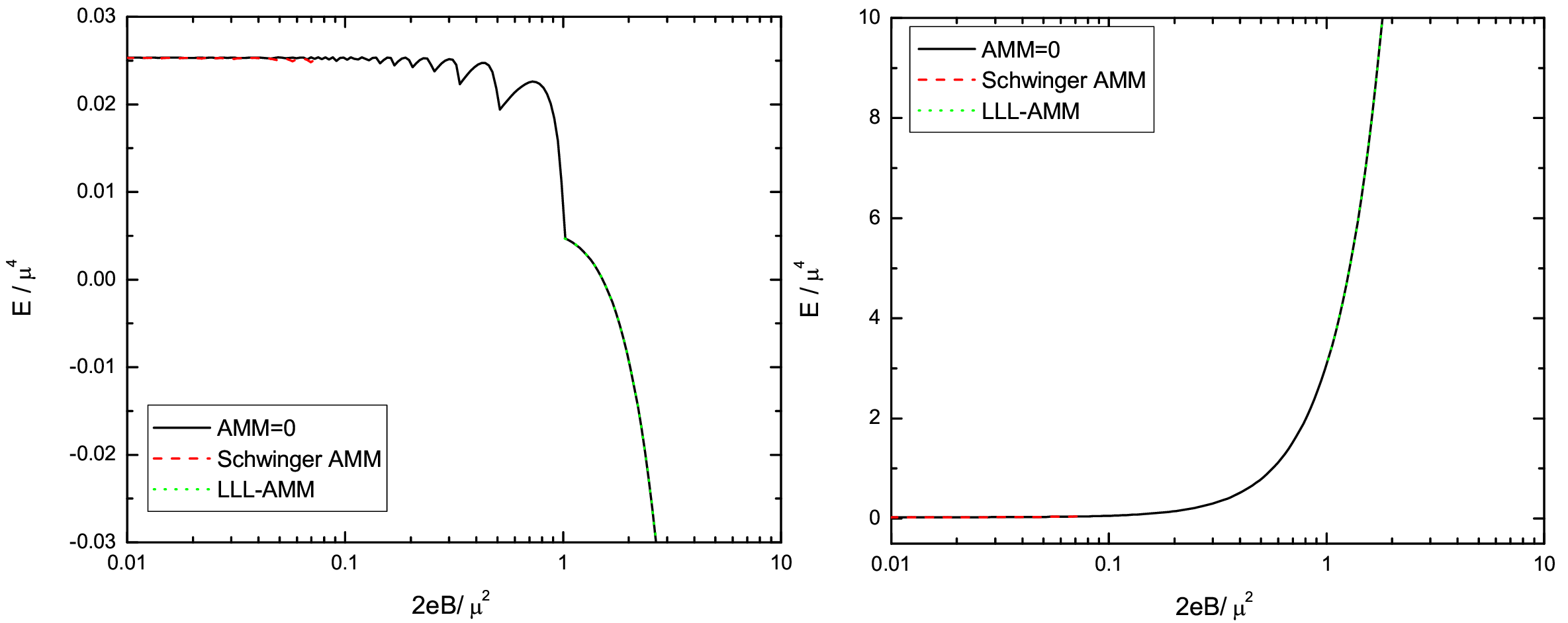}
\caption{\footnotesize Energy density versus $2eB/\mu^2$, with $m/\mu\simeq 0.02$. Left panel corresponds to energy density without the Maxwell contribution $(B^2/2)$; on the right panel the Maxwell contribution has been included. Comparison between AMM in LLL, Schwinger approximation of AMM, and  without AMM, cases. } \label{fig3}
\end{center}
\end{figure}
In Figs. (\ref{fig2}),  (\ref{fig3}) and (\ref{fig4}) we have plotted $\overline{\mathcal{M}}$,  $\overline{E}$, $\overline{P}_\|$, and $\overline{P}_\bot$, versus $2eB/\mu^2$ for three cases: a) in the strong-field region, considering the contribution of the AMM obtained in (\ref{LLLMT2}), b) at weak fields, considering the Schwinger AMM, and c) at arbitrary field with no AMM.
\begin{figure}[!ht]
\begin{center}
\includegraphics[width=14cm,height=6cm]{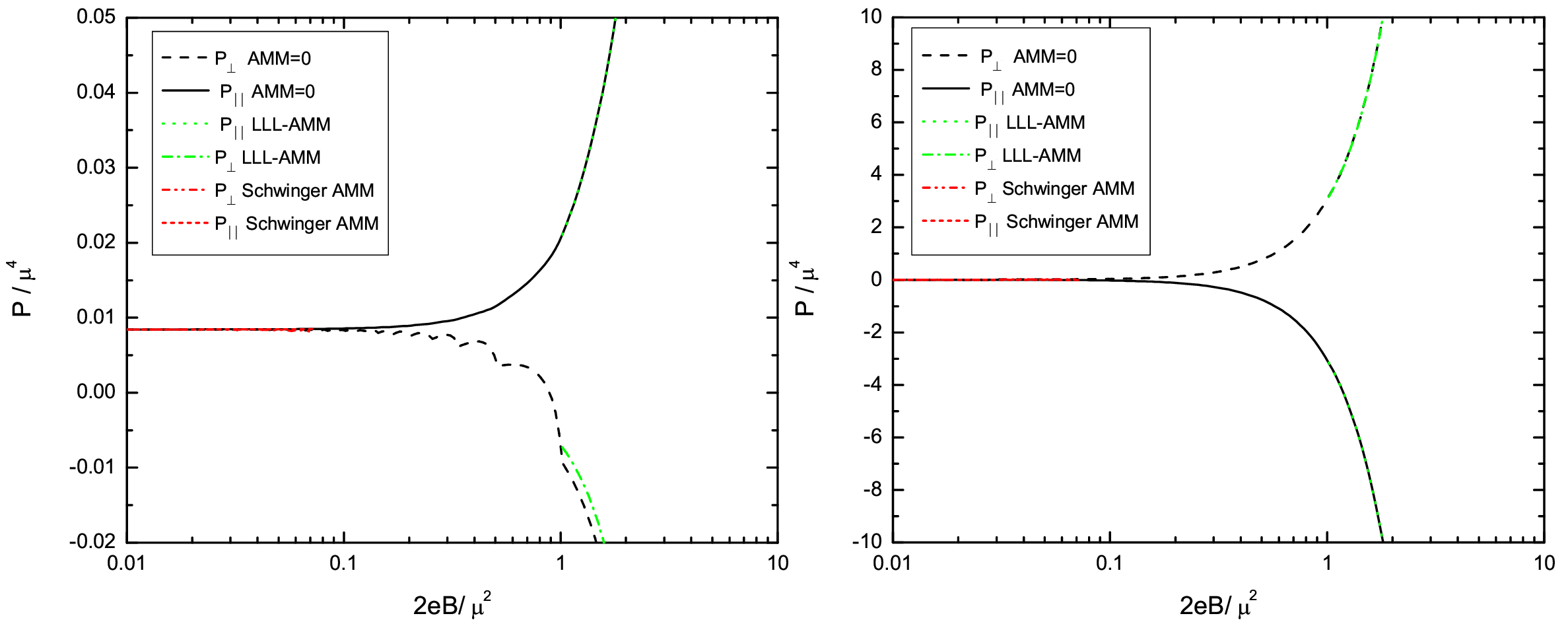}
\caption{\footnotesize Pressures parallel and perpendicular versus $2eB/\mu^2$, with $m/\mu\simeq 0.02$. Left panel corresponds to pressures without the Maxwell contribution $(B^2/2)$ and right panel with the Maxwell contribution included.  Comparison between AMM in LLL, Schwinger approximation of AMM, and  without AMM, cases. } \label{fig4}
\end{center}
\end{figure}
Fig. (\ref{fig2}) shows that the magnetization calculated with and without the inclusion of the AMM  has only slightly differences at weak and strong field. The curves without AMM and with the Schwinger AMM exhibit the Haas-van Alphen oscillations at weak field, due to the change in Landau levels.

In Figs. (\ref{fig3})-(\ref{fig4}) we have depicted  the  curves for the energy density and pressures versus $2eB/\mu^2$ respectively. In the left panels without taking into account the Maxwell contribution and right panels with the Maxwell terms already included. In the left panels we can observe the Haas-van Alphen oscillations at weak field for the energy density and transverse pressures.
The figures show that the energy density and pressures with and without the inclusion of the AMM  have only slight differences in the whole field range.


\section{Conclusions}
We presented the calculations of the quantum corrections of the AMM for  fermions in the presence of a strong magnetic field  using the Ritus's approach, which allows us to diagonalize the self-energy in momentum space and separate its LLL contribution.

We found that at strong fields the particles get different AMM's depending on the LL's. This result is different from what is obtained with the Schwinger's approximation at weak field where the AMM is independent of the LL.

We investigated how the obtained AMM affects the thermodynamics of the strongly magnetized system, and found that its effect is negligibly small. This result contradicts certain claims in the literature \cite{EoS-AMM2} about significant effects of the AMM at strong fields.

We also compared the effect of the AMM in the EoS (energy density and pressures) of the system in the weak-field regime using the Schwinger's approximation. We got a similar outcome to the strong-field case: No significant variation with respect to the zero AMM case.

\bigskip
E.J.F, V.I and A.P.M  acknowledges  the  Organizer Committee of the CSQCD3 for their support and hospitality in Guaruja, Sao Paulo, Brazil. A.P.M and D.M.P A.P.M. thanks to A. Cabo and H. Perez Rojas for discussions. The work of A.P.M and D.M.P. have been supported  under the grant CB0407 and the ICTP Office of External Activities through NET-35.

\end{document}